\documentclass[preprint,showpacs,preprintnumbers,amsmath,amssymb]{revtex4}
\usepackage{graphicx}
\usepackage{dcolumn}
\usepackage{bm}

\begin{document}

\title[Pressure effects on carbon nanotubes]{Hydrostatic pressure effects on the structural and electronic properties of carbon nanotubes}

\author{Rodrigo B. Capaz$^{1,2,3}$, Catalin D. Spataru$^{2,3}$, Paul Tangney$^{2,3}$, 
Marvin L. Cohen$^{2,3}$, and Steven G. Louie$^{2,3}$}

\affiliation{$^1$ Instituto de F\'\i sica, Universidade Federal do Rio de Janeiro, Caixa Postal 68528, Rio de Janeiro, RJ 21941-972, Brazil \\
$^2$ Department of Physics, University of California at Berkeley, Berkeley, CA 94720 \\
$^3$ Materials Science Division, Lawrence Berkeley National Laboratory, Berkeley, CA 94720
}

\begin{abstract}
We study the structural and electronic properties of isolated
single-wall carbon nanotubes (SWNTs) under hydrostatic pressure using a
combination of theoretical techniques: Continuum elasticity models,
classical molecular dynamics simulations, tight-binding electronic
structure methods, and first-principles total energy calculations within
the density-functional and pseudopotential frameworks. For
pressures below a certain critical pressure $P_c$, the SWNTs' structure
remains cylindrical and the Kohn-Sham energy gaps of semiconducting
SWNTs have either positive or negative pressure coefficients depending
on the value of $(n,m)$, with a distinct "family" (of the same $n-m$) behavior. The
diameter and chirality dependence of the pressure coefficients can be
described by a simple analytical expression. At $P_c$, molecular-dynamics
simulations predict that isolated SWNTs undergo a pressure-induced
symmetry-breaking transformation from a cylindrical shape to a collapsed geometry. This
transition is described by a simple elastic model as arising from the
competition between the bond-bending and $PV$ terms in the enthalpy.
The good agreement between
calculated and experimental values of $P_c$ provides
a strong support to the ``collapse'' interpretation of the experimental transitions in bundles.

\end{abstract}


\maketitle                   

\section{Introduction}

Since their discovery \cite{iijima}, carbon nanotubes have been the subject of
extensive investigation. Their variable electronic and mechanical
properties \cite{dresselhaus} make them ideal candidates for a variety of applications \cite{baughman}. 
Most studies of hydrostatic pressure effects in these materials have focused
on the analysis of structural changes in bundles \cite{chesnokov,venka1,peters,tang,sharma,rols,sluiter,sandler,venka2,elliott,reich}. 
The interpretation of these changes (ovalization, collapse or polygonization) seems to be 
obscured by the polydispersity of nanotube diameters in bundles \cite{sluiter,elliott}. Therefore,
the study of the mechanical response of a single, isolated single-wall carbon nanotube
(SWNT) under hydrostatic pressure is of significant value. 

Also of great importance is the study of changes in the electronic structure of SWNTs under
hydrostatic pressure. Changes in the gap of semiconducting SWNTs upon different types of 
externally applied deformations have been considered theoretically by many authors \cite{charlier,kane,heyd,yang1,lammert,yang2,mazzoni,gulseren,gartstein}, but so far the case of hydrostatic pressure has not 
been addressed. As a special motivation, recent experiments \cite{wu} in sodium dodecylsulfate (SDS)-wrapped 
SWNTs suspended in D$_2$O show a reduction in the optical gap of all measured SWNTs (with varying diameters 
and chiralities) upon increasing  pressure, a remarkable result that calls for theoretical interpretation.
 
In this work, we address the changes in structural and electronic properties of 
semiconducting SWNTs under hydrostatic pressure. Section 2 describes our methodology, consisting of a hierarchy
of theoretical techniques, ranging from continuum elasticity models and classical molecular dynamics 
simulations to tight-binding electronic structure methods and first-principles total energy calculations.
In Section 3, structural and electronic properties in the low-pressure regime (below a critical pressure 
$P_c$) are presented. In Section 4, the collapse transition at $P_c$ is described. 
Section 5 lists our main conclusions.

\section{Methodology}

Our first-principles calculations are based on density-functional theory (DFT) \cite{dft} 
within the local density approximation (LDA) with the Perdew-Zunger exchange-correlation functional
\cite{pz}. {\it Ab initio} Troullier-Martins pseudopotentials \cite{tm} are used. Calculations
are performed using the SIESTA code \cite{siesta}, which expands the Kohn-Sham wavefunctions in a 
linear combination of atomic orbitals (LCAO). A double-zeta plus polarization
(DZP) basis is used. The grid cutoff representing the charge density corresponds to a plane-wave energy 
cutoff of 240 Ry. Isolated nanotubes are modeled by a periodic supercell with sufficiently large
lateral dimensions. The irreducible Brillouin Zone is sampled with a 1x1x8 Monkhorst-Pack grid \cite{mp}. 
The effects of hydrostatic pressure $P$ in a SWNT are simulated by imposing a constraint
of radially directed external forces with magnitude $PA$, where $A$ is the surface area per
carbon atom. The surface area is calculated by using the {\it external} radius of the nanotube, 
i.e. the average nanotube radius plus a van der Waals ``exclusion distance'' of $\Delta r = 1.675$ 
\AA { }  (half of the graphite interlayer distance) \cite{venka1}. The unit cell length along the
axis ($z$ direction) is adjusted so that the $zz$ component of the stress tensor (scaled by the
ratio between the supercell and nanotube cross-sectional areas) matches the target
pressure $P$. This scheme provides an accurate modeling of structural and electronic properties (within
a single-particle description) of an individual carbon nanotube
under hydrostatic pressure at $T$ = 0 K. We apply it to several zig-zag SWNTs in the
5.5 \AA { } to 15 \AA { } diameter range.  

First-principles calculations are more time-consuming for chiral tubes. However, structural 
properties are not strongly dependent on chirality. Therefore 
one can combine the elastic constants calculated within first-principles for zig-zag tubes 
with an empirical tight-binding (TB) model to calculate the electronic properties of chiral
tubes. Although less reliable, the TB calculations are very useful for studying chirality trends
since theyr are computationally less demanding.
We use an orthogonal basis of  
one $p$ orbital per carbon atom, with nearest-neighbor hopping matrix element $\gamma = -2.89$ eV 
for undistorted bonds \cite{antonio}. Under hydrostatic pressure, bond components along axial and circumferential 
directions are distorted according to the radial and axial elastic constants 
obtained from first-principles. Hopping matrix elements of distorted bonds are modified
according to Harrison's inverse-square rule \cite{harrison}.

Molecular dynamics (MD) simulations allow for the treatment of finite-temperature
effects and for a more realistic description of hydrostatic pressure by adding a 
pressure-transmitting medium. However, they are limited in accuracy owing to the use of classical 
interatomic potentials. We use a relatively well-tested and reliable potential, namely the 
extended Tersoff-Brenner potential \cite{brenner} to model 
the carbon-carbon bonding of the nanotubes. Hydrostatic pressure is applied to 
the nanotube using a method similar to that of Martonak {\em et al.}\cite{molteni}:
The nanotube is immersed in a highly diffusive medium of particles interacting
via a repulsive $1/r^{12}$ potential. 
The pressure during each simulation is calculated in a box within the full simulation 
cell which contains only the pressure medium by computing the contributions to the stress 
tensor from the internal and thermal energies on the particles within this box.
This procedure allows hydrostaticity to be easily monitored and alerts one to
any problems which may arise from e.g. crystallization or vitrification of the medium.

\section{Low-pressure regime}

\subsection{Structural properties}

\begin{figure}[htb]
\includegraphics[width=.6\textwidth]{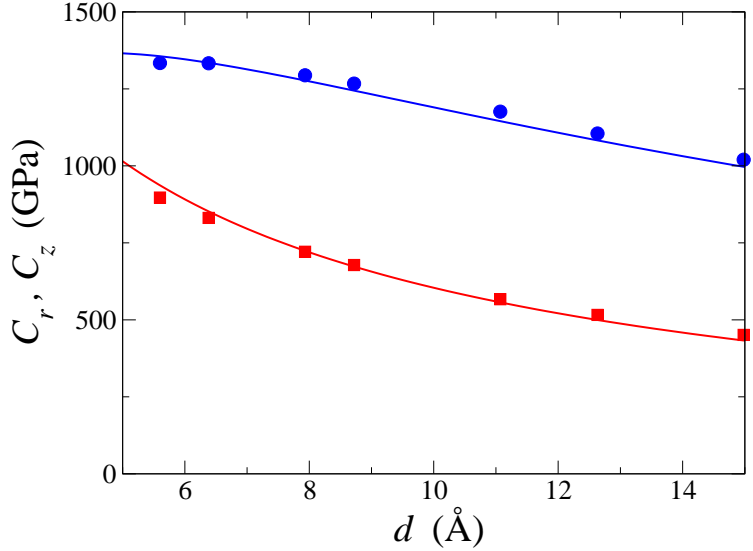}
\caption{Elastic constants $C_r$ (red squares) and $C_z$
(blue circles). Lines are best fits using expressions (\ref{eq:1}) and (\ref{eq:2}).} 
\label{fig:1}
\end{figure}

For low enough pressures (below $P_c$), SWNTs remain cylindrical in shape. Therefore, structural changes
under pressure are simply described by radial and axial strains, $\epsilon_r$ and 
$\epsilon_z$, respectively. These strains are related to $P$ by radial and
axial elastic constants, $C_r$ and $C_z$ \cite{reich}, defined as $C_r=-P/\epsilon_r$ and 
$C_z=-P/\epsilon_z$. Fig. \ref{fig:1} shows $C_r$ (squares) and $C_z$ (circles) calculated by 
first-principles as a function of $d$ for a few zig-zag tubes: (7,0), (8,0), (10,0), (11,0), (14,0),
(16,0) and (19,0). As expected for a highly anisotropic material, $C_r$ and $C_z$ are not equal. 
The same trend was obtained in  first-principle calculations in bundles \cite{reich}. Moreover, 
notice the strong diameter dependence of both elastic constants.
This dependence can be modeled by treating the nanotube as an elastic sheet with compliance constants
$S_{11}$ and $S_{12}$. By using the standard definitions of Young's modulus $Y=1/S_{11}$ and Poisson's
ratio $\nu = -S_{12}/S_{11}$, and applying geometrical considerations, we arrive at the following 
expressions \cite{elsewhere}: 
\begin{equation}
\label{eq:1}
C_r = {\frac {4Yr\Delta r}{R_o(2r-\nu R_o)}}
\end{equation}
\begin{equation}
\label{eq:2}
C_z = {\frac {4Yr\Delta r}{R_o(R_o-2\nu r)}},
\end{equation}
where $R_o$ is the nanotube external radius, $R_o=r+\Delta r$. Best fits give the parameters $Y=1010$ GPa
and $\nu = 0.242$ \cite{footnote}. The agreement between the first-principles results and the elastic
model is remarkable, even for small-diameter tubes.
\bigskip
\bigskip
\bigskip

\subsection{Electronic properties}

\begin{figure}[htb]
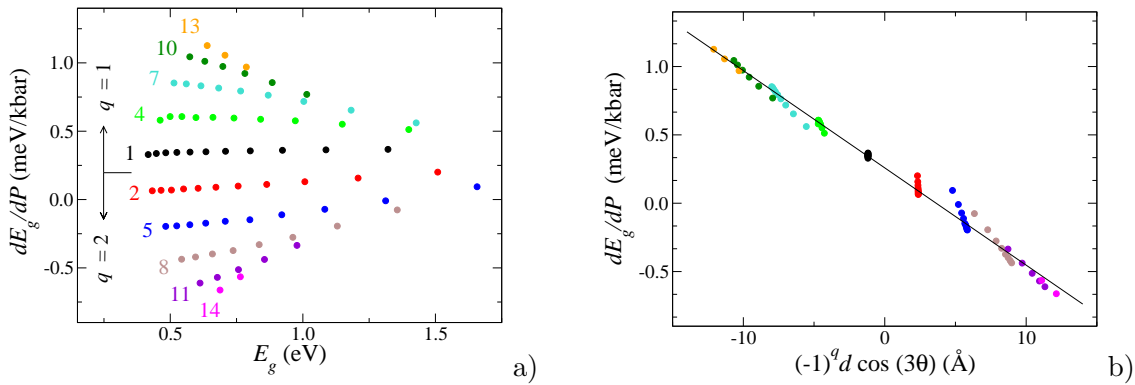

\includegraphics[width=.40\textwidth]{fig2a.eps}~a)
\hfil
\includegraphics[width=.40\textwidth]{fig2b.eps}~b)
\caption{a) Band-gap pressure coefficient as a function of the gap for a large number of
semiconducting SWNTs. Tubes are grouped into colors according to their $(n-m)$ family. The
values of $(n-m)$ for each family are also shown in the figure. b) Collapse of $dE_g/dP$ values
to a single line when plotted against $(-1)^qd cos (3\theta)$, using the same color scheme as in Fig.\ref{fig:2}(a).}
\label{fig:2}
\end{figure}

Table \ref{tab:1} shows the calculated pressure coefficients of the LDA band gap, $dE_g/dP$,  
for several zig-zag SWNTs. The data reveal two interesting features: (i) The sign of
$dE_g/dP$ depends on $q = (n-m)$ mod 3, being negative for $q=2$ and positive for $q=1$; (ii) The
magnitude of $dE_g/dP$ seems to increase with diameter. Also shown in Table \ref{tab:1} are
the values of $dE_g/dP$ calculated within TB for the same tubes, so as to compare the accuracy
of the TB model for this quantity. The overall agreement is fairly good. The simplicity of the TB
model allows us to use it on a wide range of diameters and chiralities. The results are shown 
in Fig. \ref{fig:2}(a). Each dot corresponds to a particular nanotube, and the calculated $dE_g/dP$
are plotted as a function of $E_g$ itself. A more complete picture emerges: Values
of $dE_g/dP$ seem to follow trends according to the specific values of $(n-m)$, a
so-called ``family behavior''. As a guide to the eye, results for nanotubes in the same $(n-m)$ 
family are grouped by color. The $q$-dependent sign oscillation found for zig-zag
tubes within LDA is a manifestation of this family behavior.

\begin{table}[htb]
\caption{Pressure coefficients for LDA and TB gaps of a few zig-zag SWNTs.}
\label{tab:1}\renewcommand{\arraystretch}{1.5}
\begin{tabular}{lll} \hline
$(n,m)$ & $(dE_g/dP)_{LDA}$ & $(dE_g/dP)_{TB}$  \\ 
        & (meV/kbar)        & (meV/kbar)      \\  \hline
(7,0) & 0.63 & 0.56 \\
(8,0) & -0.17 & -0.08  \\
(10,0)& 0.80 & 0.77  \\
(11,0)& -0.40 & -0.34  \\
(14,0)& -0.60 & -0.57 \\
(16,0)& 1.10 & 1.17  \\
(19,0)& 1.20 & 1.36 \\ \hline
\end{tabular}
\end{table}

\bigskip
\bigskip
\bigskip
\bigskip
\bigskip

Similar types of family behaviors have been found, for instance, in the study of band-gap changes in SWNTs 
under uniaxial stress \cite{yang1, yang2, gartstein}. In fact, the two situations (uniaxial stress
and hydrostatic pressure) are conceptually identical, differing only in the signs and
magnitudes of the resulting radial and axial strains. Therefore, one can readily adapt the
relation for the gap shift from Gartstein {\it et al.} \cite{gartstein} to the case of hydrostatic pressure:

\begin{equation}
\label{eq:3}
\delta E_g = -{\frac {4|\gamma| a_{C-C}}{d}}\epsilon_r + 3|\gamma|(\epsilon_r-\epsilon_z)(-1)^q cos (3\theta) ,
\end{equation}
where $a_{C-C}$ is the C-C bond length and $\theta$ is the chiral angle. The pressure coefficient
then becomes:
\begin{equation}
\label{eq:4}
{\frac {dE_g}{dP}} = {\frac {4|\gamma| a_{C-C}}{C_rd}} - 3|\gamma|\left({\frac {1}{C_r}}-{\frac {1}{C_z}}\right)(-1)^q cos (3\theta).
\end{equation}
One can obtain a simpler and more useful (albeit approximate) expression by taking 
the large-diameter limit in the expressions for $C_r$ and $C_z$. In this limit, both elastic 
constants decay as $1/d$, and the resulting pressure coefficient is:
\begin{equation}
\label{eq:5}
{\frac {dE_g}{dP}} = {\frac {|\gamma|}{2Y\Delta r}}\left[a_{C-C}(2-\nu) - {\frac {3}{4}}(1+\nu)(-1)^q d cos (3\theta)\right].
\end{equation}
We test this analytical expression against our numerical results by plotting $dE_g/dP$ as a function
of $(-1)^q d cos(3\theta)$ in Fig.\ref{fig:2}(b). The data collapse into a straight line is excellent.

Equation (\ref{eq:5}) has a very clear physical meaning. Hydrostatic pressure causes an overall shortening
of C-C bonds, therefore increasing, in a TB picture,  the magnitude of hopping matrix elements.
In a graphene sheet, this would lead to an increase in the Fermi velocity. Because, in a simplified
picture, energy gaps of semiconducting SWNTs are obtained by slicing the graphene bands, this would lead
to an overall tendency of $E_g$ to increase upon applying pressure, for all SWNTs. This is the meaning of the 
first term in Eq.(\ref{eq:5}), a chirality-independent positive constant. In addition to that, hydrostatic 
pressure breaks the triangular lattice symmetry of the parent graphene sheet due to
the difference in radial and axial strains. This leads to a relative shift of the slicing planes
with respect to the graphene Fermi point. This shift depends on chirality: For $q$ = 1 (2), the planes move
away from (closer to) the Fermi point, therefore increasing (decreasing) the gap \cite{yang2}. That is
the meaning of the second, chirality-dependent term in Eq. (\ref{eq:5}).

Experimentally, however, optical gaps of SDS-wrapped SWNTs in aqueous solution show an overall 
{\it decrease} with pressure, with magnitudes of $dE_g/dP$ almost ten times higher than our calculated
values for similar diameters \cite{wu}. A family behavior different from the one described in Eq.(\ref{eq:5})
is found. Although excitonic effects (not included here) are crucial for a quantitative description of such 
optical experiments \cite{catalin}, trends are often well described by a single-particle picture. Therefore, 
this {\it qualitative} disagreement between theory and experiment is puzzling and it should motivate
further work. It is possible that the interaction between the SDS micelles and the SWNTs plays an 
important role. Very recent experiments \cite{nicholas} in which strain is applied to SWNTs
by both differential thermal contraction upon freezing the D$_2$O solution and by hydration of a wrapping polymer
provide good qualitative agreement (positive and negative shifts depending on chirality) with our theoretical
predictions.

\begin{figure}[htb]
\includegraphics[width=.47\textwidth]{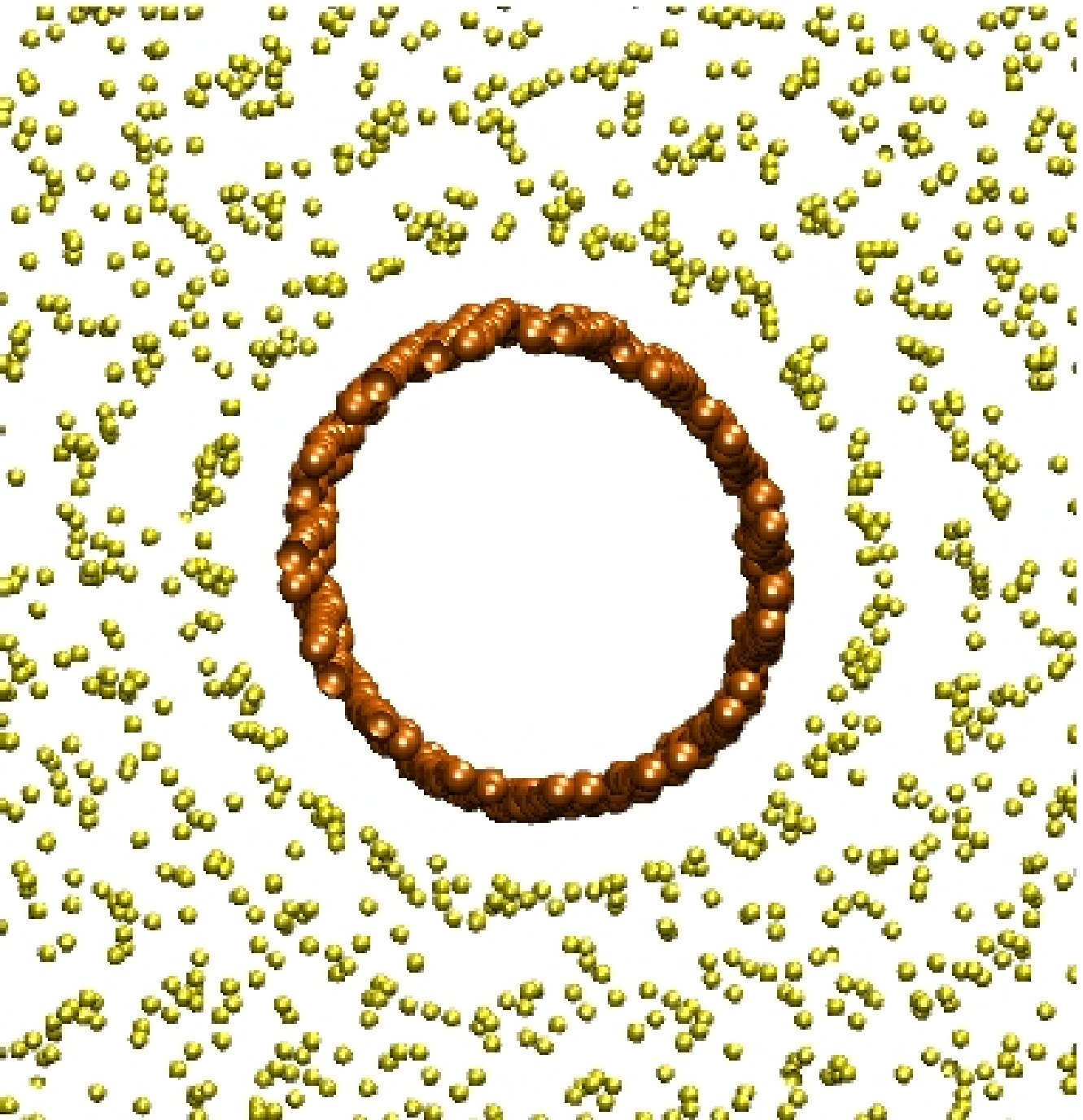}~a)
\hfil
\includegraphics[width=.47\textwidth]{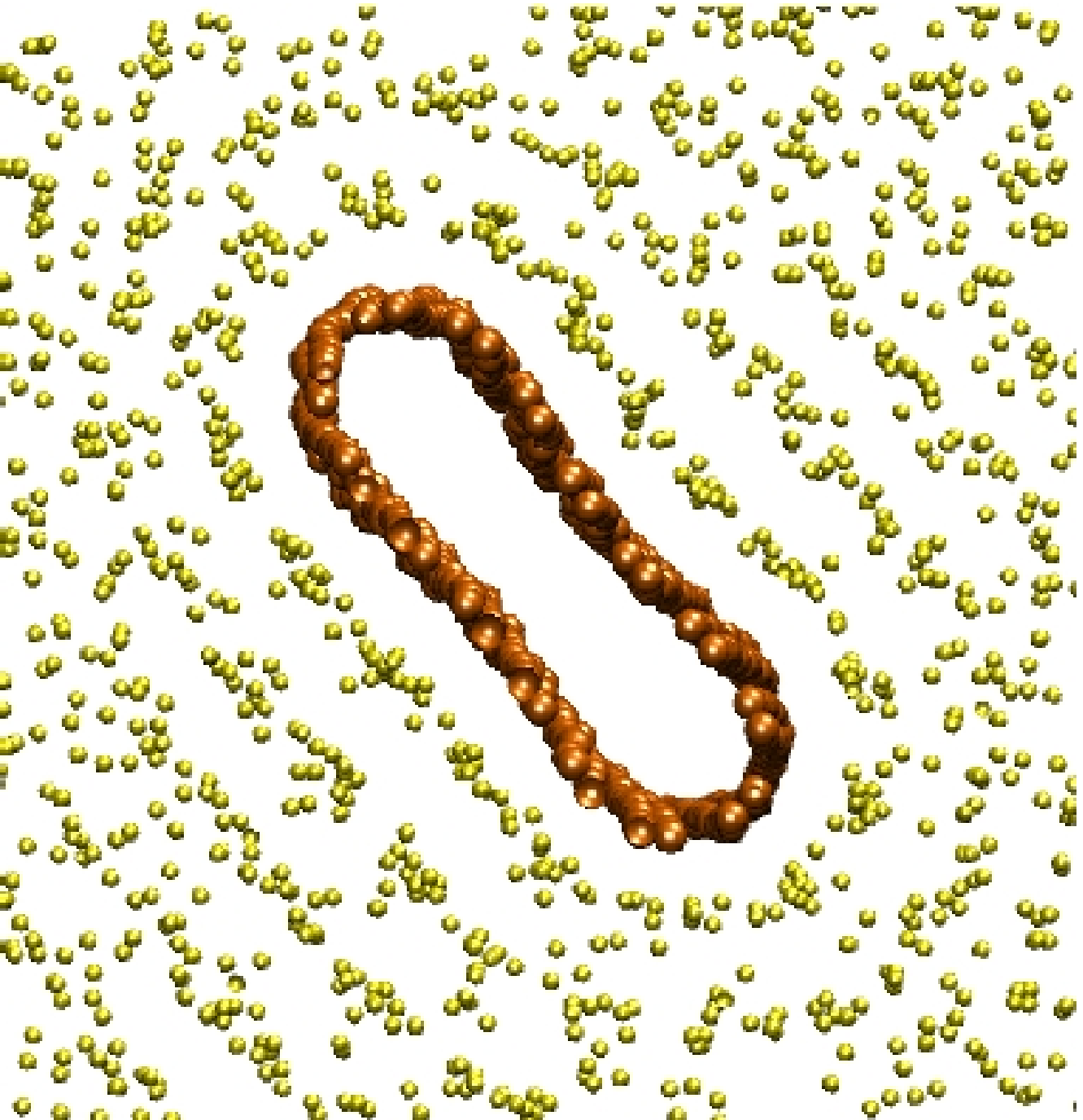}~b)
\caption{Snapshots of room-temperature MD simulations of a (8,7) SWNT at (a) 3 GPa and (b) 4GPa.
Large orange balls and small yellow balls represent carbon atoms in the nanotube and repulsive Lennard-Jones
medium particles, respectively.}
\label{fig:3}
\end{figure}

\section{The collapse transition}

Several high-pressure experiments in bundles indicate some sort of symmetry-breaking phase transition at critical
pressures ranging from 1.5 to 2.1 GPa for laser-grown tubes ($d\sim12-14${ }\AA) \cite{venka1,peters,sandler} and
at 6.6 GPa for HiPCo tubes ($d\sim8${ } \AA) \cite{elliott}. One may ask the question: Are those symmetry-breaking
transitions induced by bundling or are they related to an intrinsic property of the nanotubes? To investigate how 
{\it isolated} SWNTs respond to pressure, MD simulations are performed as described in Section 2. Figs. \ref{fig:3}(a) 
and \ref{fig:3}(b) show snapshots of room-temperature simulations at 3 GPa and 4 GPa, respectively, for of a chiral 
(8,7) SWNT ($d$ = 10.3 \AA). At 3 GPa, the tube is cylindrical, and at 4 GPa it is collapsed into a flat shape.
Therefore, isolated nanotubes indeed suffer a symmetry-breaking transition under hydrostatic pressures. Similar
conclusions have been reached by molecular mechanics \cite{li} and elastic models \cite{zang}.
In fact, Zang {\it et al.} \cite{zang} suggested that a continuous change of cross-section shape (from circular to oval
to ``peanut'' or collapsed) occurs upon increasing pressure beyond $P_c$. We do not observe any stable oval shape in 
our MD simulations.  

\begin{figure}[htb]
\includegraphics[width=.7\textwidth]{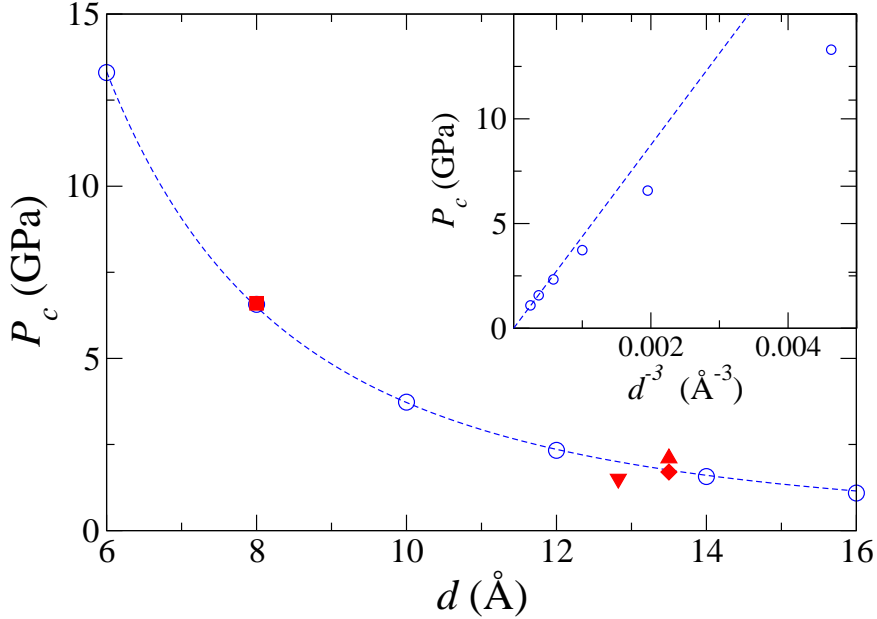}
\caption{Critical pressure for collapsing as a function of nanotube diameter. Open blue dots are
the elastic-model calculated results, the blue dashed line is a guide to the eye. Full red symbols are experimental
results. Square: Ref. \cite{elliott}, up-triangle: Ref. \cite{sandler}, down-triangle: Ref. 
\cite{venka1}, diamond: Ref. \cite{peters}. In the inset, $P_c$ is plotted against $1/d^3$. The straight line
represents the limiting $1/d^3$ behavior.} 
\label{fig:4}
\end{figure}

We also formulate a continuum elasticity model to investigate the diameter dependence of the critical
pressure $P_c$. The total enthalpy {\it per unit length} of the nanotube is written as:
\begin{equation}
\label{eq:6}
 h = {\frac {\alpha}{2C_0}}(C-C_0)^2 + \oint_C ds {\frac {\beta}{R^2}}+P{\cal A},
\end{equation}
where $C$ is the perimeter around the tube, $C_0$ is its value at $P=0$, $R$ is the local curvature
radius and ${\cal A}$ is the nanotube cross-sectional area (including the contribution from the  
van der Waals radius $\Delta r$). The first and second terms are continuum versions of the bond-stretching and 
bond-bending energies, respectively. The parameter $\alpha$ is estimated from the elasticity of graphite:
$\alpha = C_{11}c/2 =$ 3350 GPa.\AA, where $c/2$ is the interplanar distance of graphite. The parameter
$\beta$ is extracted from first-principles calculations of strain energy of nanotubes with varying
diameters \cite{mintmire}, yielding $\beta = 133$ GPa.\AA$^3$. The onset of instability 
of the cylindrical shape is searched by monitoring the change in enthalpy $\delta h$ when a small
circular-to-elliptical deformation is applied to the cross-section, at different pressures. For
 $P>P_c$, $\delta h$ should be negative. This procedure
is convenient since it allows the use of the well-known formulae for the ellipse perimeter, area and 
curvature radius. In accordance with Ref. \cite{zang}, we find that instability is driven by the
competition between the bond-bending energy (that tends to keep the circular cross-section) and
the $PV$ term (that tends to decrease the nanotube volume by collapsing it).

Figure \ref{fig:4} shows the calculated values of $P_c$ as a function of nanotube diameter.
As one can see, $P_c$ decreases with increasing diameter, as suggested in recent theoretical
works \cite{elliott, li, zang}. In the inset of Fig. \ref{fig:4}, we confirm that $P_c$ decays as
$1/d^3$ in the limit of large tubes, in agreement with the classic study by L\'evy on the theory
of elastic rings \cite{levy}. Also shown in the figure are the available experimental data.
The agreement is excellent, indicating that the instability of individual tubes under pressure
is most likely the driving force for the observed phase transitions in bundles.

\section{Conclusions}
We have investigated the structural and electronic properties of semiconducting SWNTs under
hydrostatic pressure. For low enough pressures, nanotubes remain cylindrical and are described
by radial and axial strains. Radial strains are larger than axial strains, a result that
is well described by the elastic properties of a graphene sheet. Pressure coefficients
for the single-particle band gap can be positive or negative, with strong family behavior, 
in disagreement with recent optical measurements in micelle-wrapped tubes in aqueous solution 
\cite{wu}. At a certain critical pressure $P_c$, isolated SWNTs collapse from cylindrical to a flat
shape. The good agreement between calculated and experimental critical pressures indicate
that such a pressure-induced instability of isolated SWNTs is the driving force for the
observed transitions in bundles.

We acknowledge useful discussions with J. Wu. RBC acknowledges financial support from the
John Simon Guggenheim Memorial Foundation and Brazilian funding agencies CNPq, FAPERJ,
Instituto de Nanoci{\^e}ncias, FUJB-UFRJ and PRONEX-MCT. Work partially supported by NSF
Grant No. DMR00-87088 and DOE Contract No. DE-AC03-76SF00098. Computer time was provided by the NSF at the
National Center for Supercomputing Applications and by the DOE at the 
Lawrence Berkeley National Laboratory (LBNL)'s NERSC center.


\begin{thebibliography}{10}
\bibitem{footnote} These parameters should not be interpreted as the Young's modulus and Poisson's ratio
of nanotubes, although their values are close to some of the reported values in the literature for these
quantities.
\end{thebibliography}
\end{document}